\documentstyle[aps,prd,preprint,floats,epsfig,tighten]{revtex}
\newcommand{\wfig}[4]{\begin{figure*}[tp]\vfill\begin{center}
\mbox{\epsfig{figure=#1,height=#2}}\caption{#3}\label{#4}
\end{center}\vfill\end{figure*}}

\begin{document}

\preprint{TSL/ISV-96-0130}
\title{Fixing the renormalisation scheme in NNLO perturbative QCD \\
using conformal limit arguments}
\author{J.~Rathsman\thanks{rathsman@tsl.uu.se}
}
\address{Dept. of Radiation Sciences, Uppsala University, 
 Box 535, S-751 21 Uppsala, Sweden}

\maketitle

\begin{abstract}
We discuss how the renormalisation scheme ambiguities in QCD can be
fixed, when two observables are related, by requiring the coefficients
in the perturbative expansion relating the two observables to have
their conformal limit values, i.e. to be independent of the
$\beta$-function of the renormalised coupling. We show how the
next-to-leading order BLM automatic scale fixing method can be extended
to next-to-next-to-leading order to fix both the renormalisation scale
and $\beta_2$ in a unique way. As an example we apply the method to 
the relation between Bjorken's sum rule and $R_{e+e-}$ and
compare with experimental data as well as other scheme fixing methods.
\end{abstract}

\narrowtext

\section{Introduction}

In perturbative QCD, observables are given by expansions
in the strong coupling $\alpha_s$,
\begin{equation}
R = \left(\frac{\alpha_s}{\pi}\right)^N
\left[ R_0+R_1\frac{\alpha_s}{\pi}+
R_2\left(\frac{\alpha_s}{\pi}\right)^2+ \ldots\right],
\end{equation}
where the coefficients $R_i$ can be calculated from the appropriate 
Feynman diagrams.  The individual terms in the series depends on the
renormalisation scheme one is using but the sum of the entire series is 
independent of the scheme according to the  renormalisation group
equation. However, when the series is truncated the result becomes
renormalisation scheme dependent. This
dependence is formally of higher order than the terms calculated in the
series but numerically the difference between different schemes can be
large. These differences give a theoretical uncertainty which in 
principle makes it impossible to make any absolute predictions since
any result can be obtained by a finite renormalisation. By going to
higher order in perturbation theory the renormalisation scheme
dependence becomes smaller but in principle the problem remains. One
can argue that it is only bad scheme choices that give `crazy' results
and that as long as one uses a `sensible' scheme the result will also
be `sensible'. The question then arises, what is a `sensible'
scheme?

The question of how to choose an appropriate renormalisation scheme in
QCD has been discussed many times. Three well-known
methods for choosing the renormalisation scheme 
are the `Effective Charge Scheme' by
Grunberg \cite{GRU84}, the `Principle of Minimum Sensitivity' by
Stevenson \cite{STE81} and `Automatic Scale Fixing' (BLM) by
Brodsky,  Lepage and Mackenzie \cite{BLM83}.  All these methods are
based on some more or less intuitive principle or set of arguments for
how a perturbative series should behave.

Of special interest here is the BLM method which fixes the scale
in next-to-leading order (NLO) using conformal limit arguments.
In a conformally invariant theory the
coupling $a=\alpha(\mu)/\pi$ is scale invariant, i.e.\
\begin{equation}
\frac{d a}{d \ln \mu}=\beta(a)=
-\beta_0a^2-\beta_1a^3-\beta_2a^4-... =0.
\end{equation}
It is therefore natural to define the 
conformal limit of perturbative QCD as the limit
$\beta_i \to 0$ \cite{BLM83,SLAC6481}. 
This means that the coefficients $R_i$ in the 
perturbative series have their conformal limit values if they
do not contain any  explicit dependence on the $\beta$-function.
For example, in NLO the perturbative coefficients
should have no explicit $\beta_0$-dependence. 
In the BLM
method this is achieved by absorbing all $\beta_0$-dependent NLO terms
($\beta_0=\frac{11}{2}-\frac{1}{3}N_f$ where $N_f=$ number of
active quark flavours)
into the running of $\alpha_s$ by a suitable redefinition of the
renormalisation scale. 
It should be noted that 
the renormalisation scale obtained by the BLM method can also be
interpreted as the mean value of the virtualities in the gluon
propagators \cite{BLM83,LEP93,SLAC6481,NEU95,BAL95}.

A useful concept when discussing renormalisation scheme
uncertainties is the effective charge \cite{GRU84} of an observable
which contains all QCD-corrections. For example, 
the effective charge $\hat{a}_R$ of $R_{e^+e^-}$ is defined by,
\begin{eqnarray}
R_{e^+e^-}(Q_R) & = &
\frac{\sigma(e^+e^-\to \mbox{hadrons})}{\sigma(e^+e^-\to \mu^+\mu^-)}
 \, = \, \nonumber \\ & = &
3\sum_{i=1}^{N_f} e_i^2\left(1+\frac{3}{4}C_F\hat{a}_R(Q_R)\right).
\label{rdef}
\end{eqnarray}
Each effective charge has its own $\beta$-function \cite{GRU84}
connected to it,
\begin{equation}\label{rgeinv}
\frac{d \hat{a}_R}{d \ln Q_R} \, = \, \hat{\beta}_{R}(\hat{a}_R) \, = \,
-\beta_0\hat{a}_R^2-\beta_1\hat{a}_R^3-\hat{\beta}_{2,R}\hat{a}_R^4- 
\ldots \, ,
\end{equation}
where $\beta_0$ and $\beta_1$ are renormalisation scheme
independent and $\hat{\beta}_{i,R}$, $i\geq2$ are renormalisation 
scheme invariants. Thus, for each physical observable $A$ 
there is a specific 
$\hat{\beta}_{2,A}$ connected to it which is an inherent property
of the effective charge.
The perturbative series for an effective charge depends on the 
renormalisation scheme even in the conformally invariant theory
but when two effective charges are related, one gets a relation that
is independent of the intermediate scheme that was used.

In this paper we present a new generalisation of the BLM
method to next-to-next-to-leading order (NNLO) using the conformal
limit arguments as starting point which fixes both the renormalisation
scale and $\beta_2$ when two physical observables are related.
The value for $\beta_2$ that is obtained is an intermediate
value between the $\hat{\beta}_2$'s of the two effective charges.
(A generalisation to the factorisation scheme
problem will be considered in a separate paper \cite{JRPREP}). 
This is a variation of an approach by
Grunberg and Kataev \cite{GRU92}, but whereas they claimed
that the prescription for making the coefficients $N_f$-independent
is ambiguous, we will show that once the initial renormalisation scheme
is fixed by relating two physical observables, the conformal limit 
arguments fixes the scheme in a unique way.
We also compare with the
single \cite{SLAC7085} and multi-scale extensions 
\cite{SLAC6481,GRU92,SUR93} of the BLM method to NNLO which fixes
the renormalisation scale when two effective charges $A$ and $B$
are related, $\hat{a}_A=\hat{a}_B(1+r_{1,A/B}\hat{a}_B+\ldots)$, 
using $\hat{\beta}_{2,B}$.
As an example the conformal limit scheme fixing method is applied
to the relation between Bjorken's sum rule in polarised deep inelastic
scattering and $R_{e+e-}$. The result is compared with a recently
reported experimental determination of Bjorken's sum rule and the
general renormalisation scheme dependence.

\section{The conformal limit scheme fixing method}

Consider an observable in NNLO depending on one energy scale $Q$ 
such as $R_{e^+e^-}(Q_R)$ defined by Eq.~(\ref{rdef}). The effective
charge $\hat{a}_R$ contains all QCD-corrections,
\begin{eqnarray}
\hat{a}_R(Q_R) \, = \,  a(\mu,\beta_2,...) 
\left[\right. & & 1+r_1(Q_R,\mu)a(\mu,\beta_2,...) +
\nonumber \\ & & +
r_2\left.(Q_R,\mu,\beta_2)a^2(\mu,\beta_2,...)\right],
\end{eqnarray}
where the coefficients $r_i$ can be calculated using perturbative QCD.
The renormalisation scheme dependence can be parametrised
through the renormalisation scale $\mu$ and the coefficients in the
$\beta$-function, $\beta_i$ for $i\geq2$
\cite{STE81}. Strictly speaking it is the ratio, $\mu/\Lambda$, 
of the renormalisation
scale and the QCD  scale parameter $\Lambda$ that is
the relevant parameter but in the following we will often make the
implicit  assumption that $\Lambda$ is held fixed when $\mu$ is varied.
This can be done by choosing a measurement of an effective charge to
define  $\Lambda$ as will be shown later.

The first two terms in the renormalisation group equation for the 
coupling $a=\alpha_s(\mu)/\pi$,
\begin{equation} \label{rengroup}
\frac{d a}{d \ln \mu}=\beta(a)=
-\beta_0a^2-\beta_1a^3-\beta_2a^4-... \, ,
\end{equation}
are renormalisation scheme independent,
\begin{eqnarray}
\beta_0 & = & \frac{11}{6}N_C-\frac{1}{3}N_f \, , \\
\beta_1 & = & \frac{17}{12}N_C^2-\frac{5}{12}N_CN_f-\frac{1}{4}C_FN_f
\, ,
\end{eqnarray}
whereas the higher order terms depend on the renormalisation scheme.

Applying self-consistency for the perturbative expansion of the
effective charge with respect to the renormalisation
scheme parameters,
\begin{eqnarray*}
\frac{d \hat{a}_R}{d \ln \mu} \, , \: 
\frac{d \hat{a}_R}{d \beta_2} & = & {\cal O}(a^{4}) \, ,
\end{eqnarray*}
gives \cite{STE81} the following renormalisation scheme invariants,
\begin{eqnarray}\label{r1inv}
\hat{r}_1 & = & r_1-\beta_0\ln\frac{\mu}{\Lambda} \, ,\\
\label{beta2inv}
\hat{\beta}_{2,R} & = & \beta_2-\beta_1r_1-\beta_0r_1^2+\beta_0r_2 \, ,
\end{eqnarray}
where $\hat{\beta}_{2,R}$ is 
the coefficient in the renormalisation group  equation for the
effective charge given by Eq.~(\ref{rgeinv}).
In passing we also note that the expression for the
renormalisation scheme invariant $\hat{r}_1$ 
shows explicitly that it is $\mu/\Lambda$ that is the relevant
parameter for parametrising the renormalisation scheme dependence. 

From the self-consistency requirements we also get the explicit 
$\mu$- and $\beta_j$-dependence of the coefficients $r_i$,
\begin{eqnarray}\label{r1dep}
r_1 & = & r_1^* + \beta_0\left[d^*+\ln\frac{\mu}{Q}\right] , \\
\label{r2dep}
r_2 & = & r_2^* - \frac{\beta_2-\beta_{2,\overline{MS}}}{\beta_0} + 
           \beta_1\left[d^*+\ln\frac{\mu}{Q}\right] + 
\nonumber \\ & &
        + \beta_0\left[e^*+2r_1\ln\frac{\mu}{Q}\right] +
          \beta_0^2\left[f^*-\ln^2\frac{\mu}{Q}\right] ,
\end{eqnarray}
where we have assumed that the coefficients have been calculated in the
$\overline{MS}$ scheme with $\mu=Q$ to fix the integration constants.
(The $^*$ is used to indicate terms that are 
independent of $\beta_0$ and $\beta_1$.)
We also assume that $r_1$ and $r_2$ only contain $\beta_0$- and
$\beta_1$-dependent terms from loop-insertions which is why the
$\beta_0$ term in $r_1$ and the $\beta_1$ term in $r_2$ are
the same, i.e\ they are both given by $d^*$. This way we also fix the
redundancy in how to divide $r_2$ into $\beta_0$ and
$\beta_1$ dependent parts.

We are now in the position to apply the conformal limit arguments to
the effective charge $\hat{a}_R$ to fix the renormalisation scheme 
parameters $\mu$ and $\beta_2$.
First the renormalisation scale is fixed by requiring $r_1$ to be
$\beta_0$-independent.
From Eq.~(\ref{r1dep}) we see that this can be obtained by choosing the 
renormalisation scale as
\begin{equation}
\mu^*=\mu_{BLM}=Q\exp(-d^*).
\end{equation}
We also note that the renormalisation scale obtained in this 
way is the same as in the original BLM method. 

Next $\beta_2$ is fixed by requiring $r_2$ to be $\beta_0$- and 
$\beta_1$-independent, i.e.\ $r_2=r_2^*$.
Using the renormalisation scheme invariant 
$\hat{\beta}_{2,R}$ we get the following expression for $r_2$,
\begin{equation}
r_2=(r_1^*)^2+ \frac{\beta_1}{\beta_0}r_1^* + 
\frac{\hat{\beta}_{2,R}-\beta_2}{\beta_0} .
\end{equation}
From this we see that by choosing a renormalisation scheme 
where $\beta_2$ is given by,
\begin{equation}\label{beta2star}
\beta_2^*= 
\hat{\beta}_{2,R}+\beta_1r_1^*+\beta_0(r_1^*)^2-\beta_0r_2^* \, ,
\end{equation}
we get $r_2=r_2^*$. Note that this value of $\beta_2$ in general  is
different  both from the effective charge value, $\hat{\beta}_{2,R}$ 
and from $\beta_{2,\overline{MS}}$ which was used in the calculation.
However, if $r_i^*=0$ then $\beta_2^*=\hat{\beta}_{2,R}$ 
and if $r_i^*=r_i$ then
$\beta_2^*=\beta_{2,\overline{MS}}$.

This fixes the renormalisation scheme in NNLO 
up to the question of initial
scheme, which is resolved when two physical observables are related
as shown below. This does not introduce any new uncertainties since
only relations between observables can be predicted in a 
renormalised theory and for each pair of observables we get a 
unique relation. The situation here is not different from  what happens
in the BLM method and its earlier extensions where it is also 
necessary to fix the initial renormalisation scheme to get a
unique result.
In \cite{GRU92} it was argued that `in QCD, setting $r_i=r_i^*$ is 
always possible, but leaves us with an ambiguous prescription'.
However, as we have shown above, there are no ambiguities once 
the initial renormalisation scheme has been fixed and this can be
done using a physical observable as shown below.

The perturbative series for the effective charge $\hat{a}_R$ in NNLO
thus becomes
\begin{eqnarray}
\hat{a}_R & = & a^*\left(1+r_1^*a^*+r_2^*(a^*)^2\right)
\end{eqnarray}
where the $r_i^*$'s contain no explicit $\beta_j$-terms.
In this way we obtain the required feature that all signs of scale 
breaking, i.e.\ $\beta \neq 0$, is  confined into the running 
of the coupling and the coefficients in the perturbative series 
have their conformal limit values.
Finally $a^*$ can be obtained by solving the renormalisation group
equation (\ref{rengroup}) with the fixed $\beta_2^*$.

Before ending this section we note that the method for fixing 
$\beta_2$ can be generalised to arbitrary order, $n \geq 2$.
For this we need the renormalisation scheme
invariants $\hat{\beta}_{n,R}$ in the renormalisation group equation
for the effective charge Eq.~(\ref{rgeinv}). The general form
for $\hat{\beta}_{n,R}$ is given in \cite{GRU84} and can be 
rewritten as,
\begin{equation}
r_n = \sum_{j=0}^nc_{j,n}r_1^n +
\frac{1}{n-1}\frac{\hat{\beta}_{n,R}-\beta_n}{\beta_0}
\end{equation}
where $c_{j,n}$ only depends on $\{\hat{\beta}_{i,R},\beta_{i}\}$ 
with $i\leq n-1$. 
In previous steps of applying the conformal limit arguments, 
the renormalisation scale
has been fixed so that $r_1=r_1^*$ and the $\beta_{i}$'s, 
($2\leq i \leq n-1$) have been
fixed to $\beta_{i}=\beta_{i}^*$.
So by requiring $r_n=r_n^*$ to
contain no explicit $\beta_{i}$-terms for $i\leq n-1$ the value of
$\beta_n$ is fixed to be,
\begin{equation}
 \beta_n^* = \hat{\beta}_{n,R} - (n-1)\beta_0r_n^* + 
 (n-1)\beta_0\sum_{j=0}^nc_{j,n}^*(r_1^*)^n 
\end{equation}
which is a generalisation of Eq.~(\ref{beta2star}) to arbitrary 
$n\geq2$.

\section{Comparison with other scale fixing methods}

In previous multi-scale extensions of the BLM method 
(denoted MBLM in the following)
\cite{SLAC6481,GRU92,SUR93} one has different scales for each 
$\alpha_s$-term, i.e.\
\begin{eqnarray}
\hat{a}(Q) & = & 
a(\mu_1)+r_1(\mu_1)a^2(\mu_2)+r_2(\mu_1,\mu_2)a^3(\mu_3)
\end{eqnarray}
where $\mu_1$ is parametrised as 
$\mu_1=\mu_0\exp[\theta a(\mu)]$
and $\mu$ as well as $\mu_3$ are arbitrary (they will be fixed in 
higher order approximations
but here we simply set them to be the same as $\mu_2$).

The MBLM scale fixing
method is constructed to have $\beta_2$ unchanged and
instead $\theta$ and
$\mu_2$ are  introduced which gives
three ($\mu_1=\mu_0\exp[\theta a(\mu)]$, $\mu_2$ and $\beta_2$) 
unphysical\footnote{when two physical observables are related 
the MBLM (and SBLM) method uses the effective charge $\hat{\beta}_2$ 
of one of the observables which in principle is a measurable quantity.}
parameters instead of the minimal two ($\mu$ and $\beta_2$).
Requiring that the effective
charge does not depend on these parameters, to the present order of 
perturbation theory, gives 
the explicit $\mu$-, $\beta_j$- and $\theta$-dependence 
of the coefficients $r_i$ ,
\begin{eqnarray}\label{MBLM1}
r_1 & = & r_1^* + \beta_0\left[d^*+\ln\frac{\mu_0}{Q}\right] , \\
\label{MBLM2}
r_2 & = & r_2^* - \frac{\beta_2-\beta_{2,\overline{MS}}}{\beta_0} 
        + \beta_1\left[d^*+\ln\frac{\mu_0}{Q}\right] + \beta_0\theta +
\nonumber \\ & &
        + \beta_0\left[e^*+2r_1\ln\frac{\mu_2}{Q}\right] +
          \beta_0^2\left[f^*-\ln^2\frac{\mu_0}{Q}\right] \, ,
\end{eqnarray}
where again the integration constants are fixed by assuming that the
calculation was made in the $\overline{MS}$ scheme with $\mu=Q$. 
Comparing with Eqs.~(\ref{r1dep},\ref{r2dep}) we see the effects
of having different renormalisation scales 
and also how the $\theta$-dependence enters.
In the MBLM scale fixing 
all $N_f$-dependent terms should be absorbed so that,
$r_{1} = r_1^*$ and
$r_{2} = r_2^*$ just as in the conformal limit scheme.
Keeping in mind that $\beta_2$ should be unchanged we
see that this can be achieved by choosing 
\begin{eqnarray*}
\beta_2 & = & \beta_{2,\overline{MS}}, \\
\mu_0 & = & Q\exp(-d^*), \\
\theta & = & \beta_0(-f^*+(d^*)^2) , \\
\mu_2 & = & Q\exp\left[-e^*/(2r_1^*)\right],
\end{eqnarray*}
so that $\mu_1=Q\exp\left\{-d^*-\beta_0[f^*-(d^*)^2]a(\mu_2)\right\}$.
From Eqs.~(\ref{MBLM1},\ref{MBLM2}) it is also easy to see that one 
only needs
a single renormalisation scale if $\theta$ is chosen appropriately.
In this single-scale extension \cite{SLAC7085} of the BLM scale fixing
method (denoted SBLM in the following) one chooses
$\mu_2=\mu_1=\mu_0\exp[\theta a(\mu)]$ where
$\mu_0 = Q\exp(-d^*)$ and 
$\theta=\beta_0(-f^*+(d^*)^2)-e^*+2r_1^*d^*$.

\section{Fixing the initial scheme with a physical observable}

Up to now we have assumed that the initial
renormalisation scheme (and thereby $\Lambda$) is fixed.
Now we will show how this can be accomplished using a physical
observable so that a unique prediction of another physical
observable can be made.
As an example we will relate $R$ defined in
Eq.~(\ref{rdef}), to $K$, Bjorken's sum rule for polarized 
deep-inelastic electroproduction \cite{BJSR}.

The effective charge for $R$ is in NNLO given by 
(in the $\overline{MS}$ scheme),
\begin{equation}\label{rprime}
\hat{a}_{R}=
a_{\overline{MS}}(1+r_1a_{\overline{MS}}+r_2a_{\overline{MS}}^2) \, ,
\end{equation}
where $r_1$ and $r_2$ can be obtained from \cite{GOR91,SUR91}. 
For Bjorken's sum rule,
one can also define an effective charge $\hat{a}_{K}$ (using
the same normalisation as in \cite{SMC95}),
\begin{eqnarray}\label{bjsumrule}
K & = &
\int_0^1dx [g_1^{ep}(x,Q^2)-g_1^{en}(x,Q^2)] 
\, = \, \nonumber \\ & = &
\frac{1}{6}\left|\frac{g_A}{g_V}\right|
\left(1-\frac{3}{4}C_F\hat{a}_{K}(Q)\right).
\end{eqnarray}
In NNLO $\hat{a}_{K}$ is given by,
\begin{equation}
\label{kbjorken}
\hat{a}_{K}=
a_{\overline{MS}}
(1+k_1a_{\overline{MS}}+k_2a_{\overline{MS}}^2) \, ,
\end{equation}
where $k_1$ and $k_2$ have been calculated in \cite{LAR91}.

Recognising that the $\overline{MS}$ scheme is only an intermediary 
scheme suited for calculations we can find a unique relation
between the two observables $\hat{a}_{{R}}$ and
$\hat{a}_{K}$. Inverting Eq.~(\ref{rprime}) for $a_{\overline{MS}}$ 
and inserting into Eq.~(\ref{kbjorken}) gives
\begin{equation}\label{aksol}
\hat{a}_{K}(Q_{K})=\hat{a}_{{R}}(Q_{R})
\left(1+c_1\hat{a}_{{R}}(Q_{R})+
   c_2\hat{a}_{{R}}^2(Q_{R})\right)
\end{equation}
where now $Q_{R}$ is the renormalisation scale.
The coefficients $c_i$, which are independent of the
intermediate scheme, are given by
\begin{eqnarray}
c_1 & = & -\frac{3}{4}C_F-\beta_0\left(\frac{7}{4}-2\zeta_3-
\ln\frac{Q_{R}}{Q_K}\right), \\    
c_2 & = & \frac{9}{16}C_F^2 - \ell
- \frac{\beta_2-\hat{\beta}_{2,R}}{\beta_0}
- \beta_1\left(\frac{7}{4}-2\zeta_3-\ln\frac{Q_{R}}{Q_K}\right) +
\nonumber \\ & &
+\beta_0 C_F\left(\frac{523}{144}+\frac{14}{3}\zeta_3-10\zeta_5\right) +
\nonumber \\ & &
-\beta_0 N_C\left(\frac{13}{36}-\frac{1}{3}\zeta_3\right)
+\beta_0 2c_1\ln\frac{Q_{R}}{Q_K} -
\nonumber \\ & &
- \beta_0^2
\left[-\frac{17}{6}+\left(\frac{35}{3}-8\zeta_3\right)\zeta_3-
\frac{\pi^2}{12}+\ln^2\frac{Q_{R}}{Q_K}\right],
\end{eqnarray}
where $\ell$ is the light-by-light term\footnote{Numerically
the light-by-light term is small: ${\ell}=-0.0376$ for $N_f=5$,
${\ell}=-0.1653$ for $N_f=4$ and ${\ell}=0$ for $N_f=3$. The 
light-by-light term is not affected by the conformal limit arguments
since it is not proportional to $\beta_0$.},
\begin{equation}\label{light}
{\ell} = 
  \frac{d^{abc}d^{abc}\left(\sum_{i=1}^{N_f} e_i \right)^2}
       {N_CC_F \sum_{i=1}^{N_f} e_i^2}
       (\frac{11}{144}-\frac{1}{6}\zeta_3).
\end{equation}
Hereby all dependence on the $\overline{MS}$ scheme has disappeared.
Effectively what we have done is to go from the $\overline{MS}$ scheme 
to the $R$-scheme, the effective charge scheme for $R$
which is the renormalisation scheme where 
$\hat{a}_{{R}}(Q_{R})$ has no perturbative corrections, i.e.
$\beta_2=\hat{\beta}_{2,R}$ and $\ln(\mu/\Lambda_R)=-\hat{r}_1/\beta_0$
as seen from Eqs.~(\ref{r1inv},\ref{beta2inv}).

Applying the conformal limit criteria so that the coefficients $c_1$
and $c_2$ take their conformal limit values,
$c_1^*=-\frac{3}{4}C_F=-1$ and $c_2^*=\frac{9}{16}C_F^2-\ell=1-\ell$, 
we get the conformal limit renormalisation scheme parameters 
$Q_{R}^*$ and $\beta_2^*$,
\begin{eqnarray} \label{csr}
Q_{R}^* & = & Q_{K}\exp\left(\frac{7}{4}-2\zeta_3\right),
\\ \label{cbr}
\beta_2^* 
& = & \hat{\beta}_{2,K}+c_1^*\beta_1+(c_1^*)^2\beta_0-c_2^*\beta_0 
\, =\, 
\nonumber \\ & = & 
\hat{\beta}_{2,K}-\beta_1+\ell\beta_0 \, ,
\end{eqnarray}
where $\beta_2^*$ is obtained from the invariant 
$\hat{\beta}_{2,K}=
\beta_2^*-c_1^*\beta_1-(c_1^*)^2\beta_0+c_2^*\beta_0$. 
Eq.~(\ref{csr}) has been called a commensurate scale relation 
\cite{SLAC6481} since
it gives the relation between the renormalisation scales 
when two physical observables are related. 
In the same sense one can call
Eq.~(\ref{cbr}) a commensurate $\beta_2$ relation since it gives
$\beta_2$ when two observables are related. The resulting value for 
$\beta_2^*$ is an intermediate value between the two effective
charge values $\hat{\beta}_{2,K}$ and $\hat{\beta}_{2,R}$. 
For a general relation between two effective charges,
$\hat{a}_A=\hat{a}_B(1+r_{1,A/B}\hat{a}_B+r_{2,A/B}\hat{a}^2_B)$,
$\beta_2^*$ depends on the conformal limit values
of the coefficients $r_{i,A/B}^*$. 
If $r_{i,A/B}^*=0$ then $\beta_2^*=\hat{\beta}_{2,A}$
and if $r_{i,A/B}^*=r_{i,A/B}$ then $\beta_2^*=\hat{\beta}_{2,B}$.
In other words the conformal limit scheme value $\beta_2^*$
`interpolates' between the two effective charge values,
$\hat{\beta}_{2,A}$ and $\hat{\beta}_{2,B}$, depending on
the conformal limit values of the coefficients.

Finally we have the conformal limit result in NNLO,
\begin{equation}
\label{clscre}
\hat{a}_{K}(Q_{K})=a_{R}^*
\left(1-a_{R}^*+(1-\ell)(a_{R}^*)^2\right) ,
\end{equation}
which relates one effective charge $(\hat{a}_{K})$ 
to another one $(a_{R}^*)$ which has been
modified in a unique way. This relation resembles the 
non-perturbative Crewther relation \cite{CRE72},
$3S=K R^\prime$,
which is derived using conformal and chiral invariance. 
It relates Adler's anomalous
constant ($S$), Bjorken's sum rule for polarized deep-inelastic
electroproduction ($K$) and the isovector part of R (${R^\prime}$). 
According to the no-renormalisation theorem for the axial anomaly
\cite{ADL69} one might think \cite{BRO93} that the perturbative 
corrections to
$K$ and $R$ cancel. This is not the case, but instead one
finds \cite{BRO93} that the combined corrections are proportional 
to the $\beta$-function,
$(1+\hat{a}_{R})(1-\hat{a}_{K})-1 \propto \beta(a)$,
if the light-by-light term is neglected. 
The generalised Crewther relation has been studied in more detail
in \cite{SLAC7085}.

The modified effective charge, 
$a_{R}^*(Q_{R}^*,\beta_2^*)$, can be obtained from the 
third order standard solution to Eq.(\ref{rengroup}),
\begin{eqnarray} \label{stdsol}
a_{R}^*(Q_{R}^*,\beta_2^*) & = &
\frac{1}{\beta_0\ln(Q_{R}^*/\Lambda_{R})}-
\frac{\beta_1\ln\ln(Q_{R}^*/\Lambda_{R})}
     {\beta_0^3\ln^2(Q_{R}^*/\Lambda_{R})} +
\nonumber \\ & &
+ \frac{\beta_1^2\ln^2\ln(Q_{R}^*/\Lambda_{R})
       -\beta_1^2\ln\ln(Q_{R}^*/\Lambda_{R})+
\beta^*_2\beta_0-\beta_1^2}
{\beta_0^5\ln^3(Q_{R}^*/\Lambda_{R})} \, ,
\end{eqnarray}
which is valid for $\ln(Q_{R}^*/\Lambda_{R}) \gg 1$. The value of
$\Lambda_{R}$ should be determined by experiment from 
$\hat{a}_{R}(Q_{R})$ with 
$\beta_2=\hat{\beta}_{2,R}$ and $Q_{R}=\sqrt{s}$
using the same solution for $\alpha_s$. (The definition
of $\Lambda$ depends on which solution that is used but sticking
to one definition/solution this presents no problem.)
In other words the effective charge $\hat{a}_{R}(Q_{R})$
gives an experimentally measurable $\Lambda_{R}$ and a well
defined starting  renormalisation scheme which is then modified into
the conformal limit scheme where the scheme parameters
are given by $Q_{R}^*$ and $\beta_2^*$.

\section{Discussion}

Fig.~\ref{fig1}(a,b) illustrates the renormalisation scheme dependence
of $\hat{a}_{K}(Q_K=50 \mbox{ GeV})$ as given by Eq.~(\ref{aksol})
using the standard solution, Eq.~(\ref{stdsol}), for $a_R$ with 
$\Lambda_{R}^{(5)}=502$ MeV. We see that for not too small
renormalisation scales the $\beta_2$-dependence dominates whereas for
smaller scales both the scale dependence and $\beta_2$-dependence is
quite strong. Since the renormalisation scheme dependence is
parametrised by the renormalisation scale $Q_{R}$ and $\beta_2$ when
$\Lambda_{R}$ is kept fixed, the whole space of schemes should in
principle be obtained by varying $Q_{R}$ and $\beta_2$. However, this
does not take into account the region of validity for
Eq.~(\ref{stdsol}). For too small renormalisation scales $Q_R$ or too
large $\beta_2$ this solution is no longer valid. 

\wfig{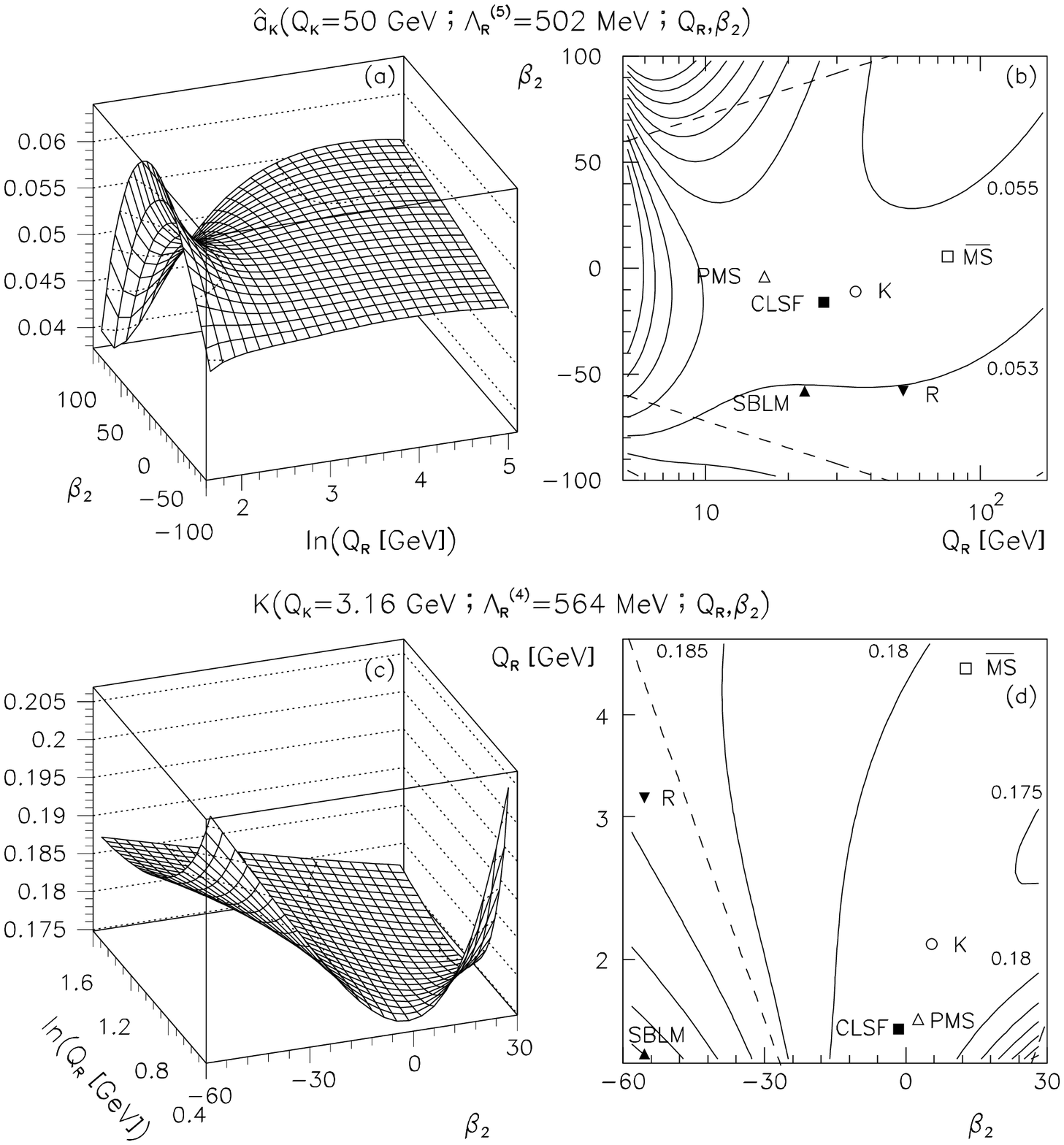}{170mm}{Renormalisation scheme dependence of 
$\hat{a}_{K}(Q_K=50 \mbox{ GeV})$ in (a,b) and Bjorken's sum rule,
$K(Q_K=3.16 \mbox{ GeV})$ in (c,d); shown as surface plots (a,c) and 
contour plots (b,d). Note that $\Lambda$ has
been kept fixed so that all the renormalisation scheme dependence is
given by the renormalisation scale $Q_{R}$ and $\beta_2$. Some
well-known scheme and scale choices are marked in (b) and (d) 
for reference and the corresponding numerical values are given in 
Table \ref{tab1} and \ref{tab2} respectively. 
The dashed lines indicates the limit of the perturbative regime, 
$|\beta_2| \leq \beta_1/a_R \simeq \beta_1\beta_0\ln(Q_R/\Lambda_R)$ 
as explained in the text. In addition to the scheme dependence 
there is also a experimental 
uncertainty from the value of $\Lambda_{R}$.}
{fig1}

To be self-consistent, one should also take into account that
Eq.~(\ref{rengroup}) has to make sense perturbatively. In other words,
if $|\beta_2| a_R \geq \beta_1$ then we are no longer in the
perturbative regime where the contribution from consecutive terms are
smaller then the preceding ones and therefore the perturbative
expansion breaks down. The lines $\beta_2 = \pm \beta_1/a_R \simeq \pm
\beta_1\beta_0\ln(Q_R/\Lambda_R)$ which indicates where this happens
are shown in Fig.~\ref{fig1}(b). These lines also indicates where the
solution given by Eq.~(\ref{stdsol}) is no longer valid and the
numerical results should therefore not be trusted in that region. 
The conformal limit scheme fixing method (CLSF) and the SBLM scale 
fixing method are indicated in Fig.~\ref{fig1}(b) together with some
other well-known schemes like the $\overline{MS}$ scheme, the
`Principle of Minimum Sensitivity' (PMS) and the effective charge
schemes (ECH) for $R$ and $K$.

Conceptually
the PMS and ECH schemes are different from the conformal limit schemes
in that they prescribe a unique scheme for each observable instead
of giving schemes for relations between observables. It should
also be noted that the PMS and ECH schemes sometimes gives 
renormalisation scales which are difficult to interpret physically. 
For instance in jet-production, both in $e^+e^-$ \cite{KRA91} and DIS
\cite{ING94}, the resulting renormalisation scales grow when the 
typical jet-mass ($y_{cut}W^2$) is decreased which is counter-intuitive.
In addition the PMS prescription depends on the intermediate
scheme. For example, applying the PMS prescription to two observables
given in the $\overline{MS}$ scheme
separately and then relating them gives a different result 
compared to if the observables are first related so that the
dependence on the $\overline{MS}$ scheme is removed and then
the PMS prescription is applied.

For reference, the numerical values of
$\beta_2$, $Q_R$, $\hat{a}_K$ and Bjorken's sum rule $K$  in the
different schemes are given in Table \ref{tab1} (together with the MBLM
method which has two renormalisation scales). Comparing the conformal
limit scheme fixing (CLSF) with the SBLM and MBLM scale fixing methods
in Table \ref{tab1} we see that even though the coefficients $c_i$ are
the same, the scales, $\beta_2$ and the resulting effective charge
$\hat{a}_K$ are different. This shows the importance of not only 
taking the commensurate scale relation into account as 
in the SBLM and MBLM methods but also the 
commensurate $\beta_2$ relation as in CLSF. 

\begin{table}[htb]
\caption{Numerical values of $\beta_2$, $Q_R$, $\hat{a}_K$ and 
Bjorken's sum rule $K$ in different schemes for $Q_K=50$ GeV and
$\Lambda_{R}^{(5)}=502$ MeV ($\Lambda$ is kept fixed so that the
scheme dependence is given by $\beta_2$ and $Q_R$). Note that there
are two scales for the MBLM method given as $Q_{R,1}(Q_{R,2})$. 
\label{tab1}}
\begin{tabular}{ddddd}
 Scheme         & $\beta_2$ & $Q_{R}$ [GeV] & $\hat{a}_{K}$ &   $K$   \\
\hline    
 CLSF           &  -15.98   &   26.00       & 0.5404        &  0.1982 \\
 SBLM           &  -57.86   &   22.18       & 0.5271        &  0.1985 \\
 MBLM           &  -57.86   &  22.17(25.02) & 0.5290        &  0.1985 \\
  $R$           &  -57.86   &   50.00       & 0.5291        &  0.1985 \\
  $K$           &  -11.00   &   33.74       & 0.5421        &  0.1982 \\
$\overline{MS}$ &    5.65   &   72.22       & 0.5444        &  0.1981 \\
 PMS            &   -3.98   &   15.88       & 0.5403        &  0.1982 \\
\end{tabular}
\end{table}

From Fig.~\ref{fig1}(b) we also see that the CLSF point is closer to
the saddle-point (PMS) than the SBLM point, which means that the scheme
dependence is smaller in the CLSF point. One might also worry that the
SBLM and MBLM scale fixing methods are too close to the line $\beta_2 =
\pm\beta_1\beta_0\ln(Q_R/\Lambda_R)$ where the perturbative expansion
for the effective charge $\hat{a}_R$ breaks down. Note that both in
Fig.~\ref{fig1} and in Table \ref{tab1} (and \ref{tab2})
the renormalisation scales are
related to $\Lambda_R$. This means that for example in the
$\overline{MS}$ scheme where one normally would use $\mu=Q_K$ as the
renormalisation scale for $\Lambda=\Lambda_{\overline{MS}}$, the scale
becomes $Q_K\exp(r_{1,\overline{MS}}/\beta_0)$ for $\Lambda=\Lambda_R$
(compare with Eq.~(\ref{r1inv})).

As a concrete example of the conformal limit scheme fixing method
we will use the global analysis of $R_{e+e-}$ data in the range 
$2.64 < Q_R <52$ GeV by Marshall \cite{MAR89} to calculate
the Bjorken sum rule $K$ at $Q_K=3.16=\sqrt{10}$ GeV,
which can be compared with the SMC measurement,
$K=\Gamma_1^p-\Gamma_1^n=0.199\pm0.038$ \cite{SMC95}.
The result of the analysis in \cite{MAR89} is a global fit 
taking both electroweak and QCD corrections into account,
$R_{e+e-}=R_{Born}^{\gamma,Z}(1+\hat{a}_R)$ ,
and numerical values for $R_Q=1+\hat{a}_R$ are given for some
distinct energies. In the following we have used the value 
$R_Q=1.0463\pm0.0044$ for $Q_R=59.2$ GeV. The reasons for picking
this particular energy are that we want to have a large scale $Q_R$ 
where the standard solution, Eq.~(\ref{stdsol}),
is a good approximation (especially since $\hat{\beta}_{2,R}$ is 
so large) and at the same time we do not want to extrapolate the
experimental result too much outside the measured range. 

Following the prescription given above for determining 
$\Lambda_R^{(5)}$ we get,
\begin{eqnarray*}
\Lambda_R^{(5)} & = & 502^{+326}_{-225} \mbox{ MeV},
\end{eqnarray*}
from $\hat{a}_R=0.0463\pm0.0044$ using Eq.~(\ref{stdsol}) with 
$\hat{\beta}_{2,R}=-57.9$ and $Q_R=59.2$ GeV.
To be able to compare with the SMC measurement
we also need $\Lambda_R^{(4)}$. This is obtained by matching $a_R^*$ 
numerically at the flavour threshold, $Q_R=m_b=5$ GeV, for $N_f=4$ 
and $5$ with $\beta_2=\beta_2^*$ which gives,
\begin{eqnarray*}
\Lambda_R^{(4)} & = & 564^{+282}_{-224} \mbox{ MeV}.
\end{eqnarray*}
The conformal limit renormalisation scheme parameters are then
obtained from Eqs.~(\ref{csr}) and (\ref{cbr}), $Q_R^*=1.64$ GeV and
$\beta_2^*=-1.56$,  and together with the
conformal limit coefficients $c_1^*=-1$, $c_2^*=1.165$ and 
$\Lambda_R^{(4)}$ this gives $\hat{a}_K =0.159^{+0.139}_{-0.048}$.
Finally the conformal limit result for Bjorken's sum rule given
by Eq.~(\ref{bjsumrule}) becomes
\begin{eqnarray*}
K & = & 0.176^{-0.029}_{+0.010} \, ,
\end{eqnarray*}
where $\left|g_A/g_V\right|=1.2573\pm0.0028$ \cite{PDG95} has been
used and the error comes from the uncertainty in $R_Q$.
This is in good agreement with the experimental value $K=0.199\pm0.038$
measured by SMC. To be able to make a more challenging test of the 
conformal limit scheme arguments one would need much more precise
measurements of both $R_{e+e-}$ and $K$.

For illustration we also show the renormalisation scheme dependence of 
Bjorken's sum rule in Fig.~\ref{fig1}(c,d) and the numerical values
of   $\beta_2$ and $Q_R$ for the specific schemes indicated in 
Fig.~\ref{fig1}(d) are given in Table \ref{tab2} together with the
resulting values for $\hat{a}_K$ and Bjorken's sum rule $K$. Comparing
with Fig.~\ref{fig1}(a,b) we see that the scheme dependence  is much
stronger which is also expected since we are at a much  smaller $Q_K$. 
We also see that the perturbative regime indicated by the dashed line 
in Fig.~\ref{fig1}(d) is narrower than in Fig.~\ref{fig1}(b) and in
fact both the SBLM and MBLM methods as well as the $R$-scheme are
outside the perturbative regime. Therefore the numerical results  given
for these schemes should not be trusted.  However, one must keep in
mind that the SBLM and MBLM scale fixing  methods advocates the use of
a physical measurement  of $\hat{a}_R$ at this scale and in this way
the problem with the validity of the solution used for $\alpha_s$ never
enters. Even so, Fig.~\ref{fig1}(c,d) illustrates clearly that there is
a strong  renormalisation scheme dependence for Bjorken's sum rule at
this scale which should be taken into account when comparing the
experimental  result with theoretical expectations.

\begin{table}[htb]
\caption{Numerical values of $\beta_2$, $Q_R$, $\hat{a}_K$ and 
Bjorken's sum rule $K$ in different schemes for $Q_K=3.16$ GeV and
$\Lambda_{R}^{(4)}=564$ MeV ($\Lambda$ is kept fixed so that the
scheme dependence is given by $\beta_2$ and $Q_R$). Note that there
are two scales for the MBLM method given as $Q_{R,1}(Q_{R,2})$. 
\label{tab2}}
\begin{tabular}{ddddd}
 Scheme         & $\beta_2$ & $Q_{R}$ [GeV] & $\hat{a}_{K}$ &   $K$   \\
\hline    
 CLSF           & -1.56     &  1.64         & 0.159	    &  0.176  \\
 SBLM           & -55.46    &  1.53         & 0.0229        &  0.205  \\
 MBLM           & -55.46    &  1.49(1.58)   & 0.0126        &  0.207  \\
  $R$           & -55.46    &  3.16         & 0.100	    &  0.189  \\
  $K$           &  5.54     &  2.09         & 0.160	    &  0.176  \\
$\overline{MS}$ & 12.70     &  4.56         & 0.147	    &  0.179  \\
 PMS            & 2.62      &  1.69         & 0.160	    &  0.176  \\
\end{tabular}
\end{table}

\section{Summary and conclusions}

In summary we have shown that it is possible to generalise the
conformal limit arguments of the BLM method to NNLO to fix the 
renormalisation scheme, i.e. both the renormalisation scale and
$\beta_2$, when two observables are related. In this way all signs of
scale breaking, i.e.\ $\beta \neq 0$, is confined into the running  of
the coupling and the coefficients in the perturbative series have their
conformal limit values.
We have also shown (contrary to \cite{GRU92}) that this prescription
for making the coefficients have their conformal limit values is
unique. Comparing with earlier extensions of the BLM method to NNLO
they only fix the scale using $\beta_2$ from the effective charge. 
This means, that the
conformal limit scheme fixing gives both a so called commensurate scale
relation as well as a commensurate $\beta_2$ relation which, in a
unique way, specifies how $\beta_2$ should be modified when two
physical observables are related to each other. 

Applying the conformal limit scheme fixing method to the relation 
between Bjorken's sum rule $K$ and $R_{e+e-}$ gives a simple relation
between the two. Using the effective charge value 
$\hat{a}_R=0.0463\pm0.0044$ for $Q_R=59.2$ GeV from
a global analysis of $R_{e+e-}$ data gives
$K=0.176^{-0.029}_{+0.010}$ for $Q^2_K=10$ GeV$^2$ where the error 
comes from the experimental uncertainty in $\hat{a}_R$. Assessing a
theoretical error is much more complicated. The theoretical 
uncertainty is illustrated by the renormalisation scheme dependence
which is shown to be quite large even though it can be reduced by
requiring the scheme to belong to the perturbative regime.  Still, the
problem of quantifying the theoretical error remains to be solved.
However, comparing with the experimentally measured $K=0.199\pm0.038$
the agreement is good and theoretical uncertainties are still smaller
than the experimental ones.

\acknowledgements
I would like to thank Stan Brodsky and Gunnar Ingelman for useful 
discussions on the subject of this paper. 
I also want to thank Hung Jung Lu for helpful
remarks on the MBLM scale fixing method.



\begin{references}
\bibitem{GRU84}
 G.~Grunberg, Phys.~Lett.~{\bf 95B}, 70 (1980); 
 Phys.~Rev.~D {\bf 29}, 2315 (1984).
\bibitem{STE81}
 P.~M.~Stevenson, Phys.~Rev.~D {\bf 23}, 2916 (1981).
\bibitem{BLM83}
 S.~J.~Brodsky, G.~P.~Lepage and P.~B.~Mackenzie, 
 Phys.~Rev.~D {\bf 28}, 228 (1983).
\bibitem{SLAC6481}
 S.~J.~Brodsky and H.~J.~Lu, Phys.~Rev.~D {\bf 51}, 3652 (1995).
\bibitem{LEP93}
 G.~P.~Lepage and P.~B.~Mackenzie, Phys.~Rev.~D {\bf 48}, 2250 (1993). 
\bibitem{NEU95}
 M.~Neubert, Phys.~Rev.~D {\bf 51}, 5924 (1995). 
\bibitem{BAL95}
 P.~Ball, M.~Beneke and V.~M.~Braun, Nucl.~Phys.~B {\bf 452}, 
 563 (1995). 
\bibitem{JRPREP}
 J.~Rathsman, TSL/ISV preprint in preparation.
\bibitem{GRU92}
 G.~Grunberg and A.~L.~Kataev, Phys.~Lett.~{\bf B279}, 352 (1992).
\bibitem{SLAC7085}
 S.~J.~Brodsky, G.~T.~Gabadadze, A.~L.~Kataev and H.~J.~Lu, 
 DESY preprint 95-245, e-Print Archive hep-ph/9512367.
\bibitem{SUR93}
 L.~R.~Surguladze and M.~A.~Samuel, Phys.~Lett.~{\bf B309 }, 157 (1993).
\bibitem{BJSR}
 J.~D.~Bjorken, Phys.~Rev.~{\bf 148}, 1467 (1966);
 Phys.~Rev.~D {\bf 1}, 1376 (1970).
\bibitem{GOR91}
 S.~G.~Gorishny, A.~L.~Kataev and S.~A.~Larin, 
 Phys.~Lett.~{\bf B259}, 144 (1991).
\bibitem{SUR91}
 L.~R.~Surguladze and M.~A.~Samuel, 
 Phys.~Rev.~Lett.~{\bf 66}, 560 (1991); {\bf 66}, 2416(E) (1991).
\bibitem{SMC95}
 SMC Coll. (D.~Adams et. al.), Phys.~Lett.~{\bf B357}, 248 (1995).
\bibitem{LAR91}
 S.~A.~Larin and J.~A.~M.~Vermaseren, 
 Phys.~Lett.~{\bf B259}, 345 (1991).
\bibitem{CRE72}
 R.~J.~Crewther, Phys.~Rev.~Lett.~{\bf 28}, 1421 (1972).
\bibitem{ADL69}
 S.~L.~Adler and W.~A.~Bardeen, Phys.~Rev.~{\bf 182}, 1517 (1969).
\bibitem{BRO93}
 D.~J.~Broadhurst and A.~L.~Kataev, Phys.~Lett.~{\bf B315}, 179 (1993).
\bibitem{KRA91}
 G.~Kramer, B.~Lampe, Z.~Phys.~A {\bf 339}, 189 (1991).
\bibitem{ING94}
 G.~Ingelman and J.~Rathsman,  Z.~Phys.~C {\bf 63}, 589 (1994).
\bibitem{MAR89}
 R.~Marshall, Z.~Phys.~C {\bf 43}, 595 (1989).
\bibitem{PDG95}
 L.~Montanet et al., Phys.~Rev.~D {\bf 50}, 1173 (1994).
\end{references}
\end{document}